\documentclass[sigconf]{acmart}

\AtBeginDocument{%
  \providecommand\BibTeX{{%
    \normalfont B\kern-0.5em{\scshape i\kern-0.25em b}\kern-0.8em\TeX}}}

\usepackage[T1]{fontenc}
\usepackage[utf8]{inputenc}
\usepackage{float}
\usepackage{comment}
\usepackage{multirow}
\usepackage{amsthm}
\usepackage{graphicx}
\usepackage{longtable}
\usepackage{tabularx}
\usepackage{changepage}
\usepackage{url}
\PassOptionsToPackage{normalem}{ulem}
\usepackage{ulem}
\usepackage{array}
\usepackage{booktabs}
\usepackage{amsmath}
\usepackage{stackrel}
\usepackage{amsmath}
\usepackage{amsthm}
\usepackage{pgfplots}
\usepackage{pgf-pie}
\usepackage{balance}
\usepackage{hyperref}
\usepackage{fontawesome5}
\usepackage{subcaption}
\usepackage{pgfplots}
\usepackage{pgfplotstable}
\pgfplotsset{compat=1.7}
\usepackage{tikz}

\hypersetup{
  colorlinks   = True, 
  urlcolor     = blue, 
  linkcolor    = blue, 
  citecolor    = blue 
}

\copyrightyear{2021}
\acmYear{2021}
\setcopyright{acmcopyright}\acmConference[ICSE-SEIP '22]{44th International Conference on Software Engineering}{May 8--27, 2022}{Pittsburgh, PA, USA}
\acmBooktitle{Software Engineering in Practice (ICSE-SEIP '22), May 8--27, 2022, Pittsburgh, USA}
\acmPrice{15.00}
\acmDOI{10.1145/xxxxxxx.xxxxxxx}
\acmISBN{978-1-4503-xxxx-x/xx/xx}

\begin{document}
\title[Decision Models for Microservices Systems]{Decision Models for Selecting Patterns and Strategies in Microservices Systems and their Evaluation by Practitioners}


\author{Muhammad Waseem$^{1}$, Peng Liang$^{1*}$, Aakash Ahmad$^{2}$\\Mojtaba Shahin$^{3}$, Arif Ali Khan$^{4}$,  Gast\'{o}n M\'{a}rquez$^{5}$}
\affiliation{%
  \institution{$^{1}$School of Computer Science, Wuhan University, Wuhan, China}
  \institution{$^{2}$College of Computer Science and Engineering, University of Ha'il, Ha'il, Saudi Arabia} 
  \institution{$^{3}$Faculty of Information Technology, Monash University, Melbourne, Australia}
  \institution{$^{4}$M3S Empirical Software Engineering Research Unit, University of Oulu, Oulu, Finland}
  \institution{$^{5}$Department of Electronics and Informatics, Federico Santa María Technical University, Concepci\'{o}n, Chile}
  \country{}}
  \renewcommand{\shortauthors}{M. Waseem et al.}

%
\begin{abstract}
Researchers and practitioners have recently proposed many Microservices Architecture (MSA) patterns and strategies covering various aspects of microservices system life cycle, such as service design and security. However, selecting and implementing these patterns and strategies can entail various challenges for microservices practitioners. To this end, this study proposes decision models for selecting patterns and strategies covering four MSA design areas: application decomposition into microservices, microservices security, microservices communication, and service discovery. We used peer-reviewed and grey literature to identify the patterns, strategies, and quality attributes for creating these decision models. To evaluate the familiarity, understandability, completeness, and usefulness of the decision models, we conducted semi-structured interviews with 24 microservices practitioners from 12 countries across five continents. Our evaluation results show that the practitioners found the decision models as an effective guide to select microservices patterns and strategies.
\end{abstract}


\maketitle       
\section{Introduction}
\label{sec:introduction}

Microservices Architecture (MSA), inspired by Service-Oriented Architecture (SOA), has gained immense popularity as an architectural style for service computing in the past few years \cite{dragoni2017microservices}. With MSAs, an application is designed as a set of business-driven microservices that can be developed, deployed, tested, and scaled independently \cite{taibi2019monolithic}. Organizations adopt MSA due to better availability, scalability, productivity, performance, fault-tolerance, and cloud support compared with SOA or monolithic applications. It is argued that MSA can also help build autonomous development teams for rapid development and delivery of software services \cite{taibi2019monolithic}.

Microservices systems entail a significant degree of complexity at the design phase and runtime configurations from an architectural perspective. Haselb{\"o}ck et al. \cite{AR5, haselbock2017decision, AR4} have identified several design areas for microservices systems, such as application decomposition, microservices security, microservices communication, and services discovery. On the other hand, literature reviews (e.g., \cite{waseemMSAdevops, waseem2020testing}), existing practices (e.g., \cite{waseemMSAdesign}), and exploratory studies (e.g., \cite{waseem2021nature}) indicate several challenges related to the design areas mentioned in \cite{AR5, haselbock2017decision, AR4}. Such issues vary from clearly defining the boundaries of microservices to security, communication, discovery, and composition aspects of MSAs.

Both academia and industry (see the identified literature in the Replication Package \cite{replpack}) have presented reusable solutions for microservices systems in the form of patterns and strategies that can help address the above mentioned challenges. These patterns and strategies are currently distributed across different publications (e.g., scientific and grey literature). The practitioners need to navigate between several patterns and strategies till a suitable combination of patterns (and strategies) is found that can address the microservices development challenge. Moreover, the practitioners cannot find a holistic view of available patterns and find themselves underprepared to select patterns and strategies and oversee their impact on Quality Attributes (QAs). According to some recent studies (e.g., \cite{waseemMSAdevops, waseem2021nature, waseem2020testing, waseemMSAdesign}), most of the design, development, monitoring and testing challenges are rooted in the design cycle of MSAs covering a multitude of aspects, including service design, deployment and discovery, (de-)composition, delivery, and security.

To assist practitioners in selecting appropriate patterns and strategies for microservices systems, we proposed the \textbf{decision models} that cover four MSA design areas: application decomposition into microservices, microservices security, microservices communication, and service discovery. Decision models are a structured way of exploring the problem and solution space to achieve the design goal(s) \cite{AR7}. In this work, the proposed models have been (1) developed by reviewing the scientific and grey literature and (2) evaluated through semi-structured interviews with microservices practitioners, which sought the practitioners' perspective on the familiarity, understandability, completeness, and usefulness of the models. The decision model for decomposing applications into microservices was proposed in our previous work \cite{waseem2021Models}. This study proposed three more decision models and evaluated all the four decision models with microservices practitioners through semi-structured interviews.

The \textbf{core contributions} of this research are: (1) four decision models that exploit patterns and strategies to accommodate quality attributes (architecturally significant requirements) for microservices systems, and (2) empirical evaluations of the decision models (accommodating practitioners perspective).

\textbf{Paper organization}: Section \ref{DecisionModel} describes the research methodology; Section \ref{sec:modelsdescripation} presents the details of the decision models; Section \ref{sec:evaluation} describes the evaluation of the decision models; Section \ref{sec:threats} discusses the threats to validity; Section \ref{sec:relatedWork} presents related work; Section \ref{sec:conclusions} concludes this work with future research directions.

\section{Methodology}
\label{DecisionModel}
\begin{figure}[!h]
 \centering
\includegraphics[width=8.5cm,height=8.5cm,keepaspectratio]{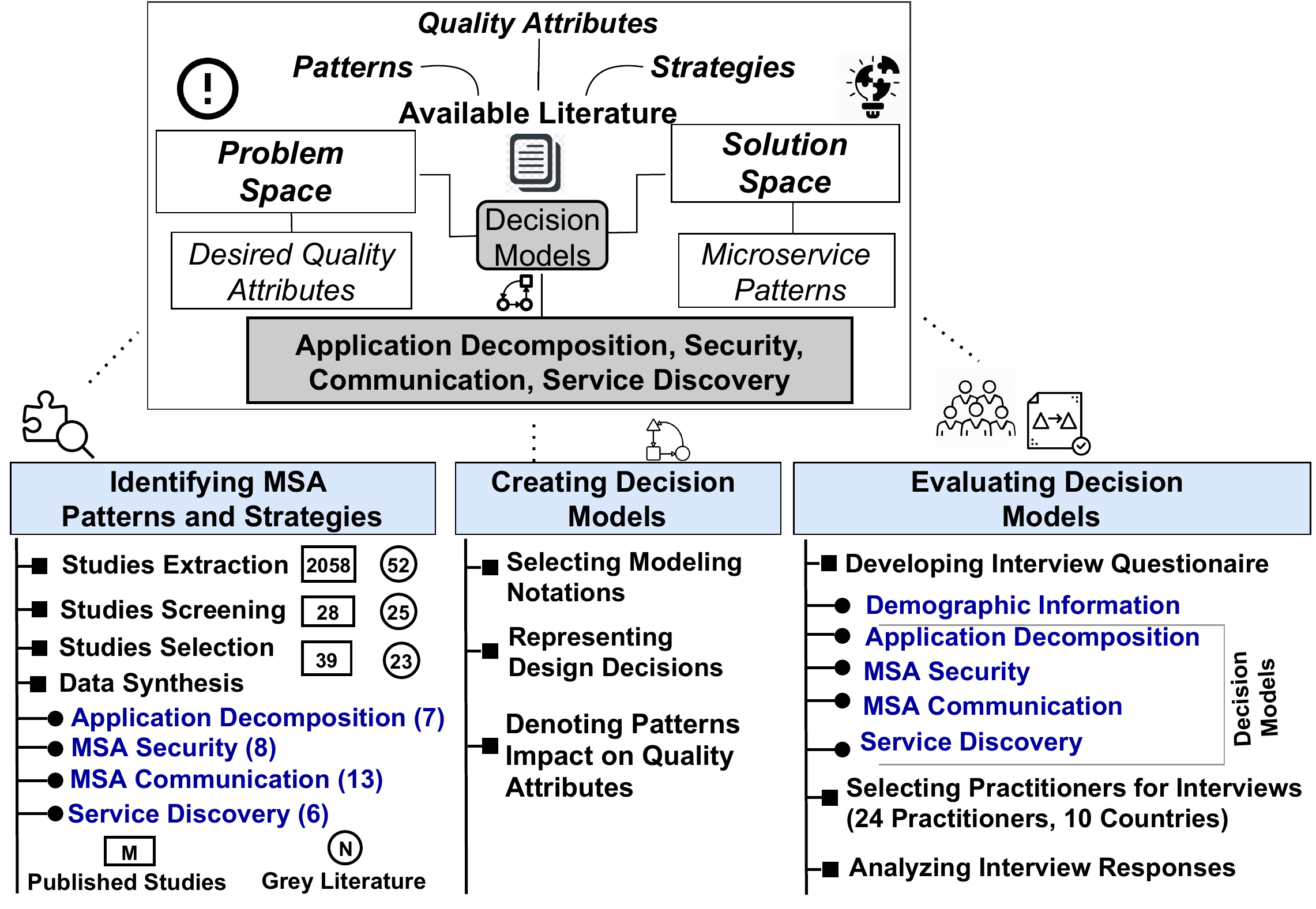}
\caption{Overview of the research methodology}
\label{fig:methodology}
\end{figure}

The decision models in software architecture are used to map elements of the problem space to elements of the solution space \cite{AR7}. The problem space represents functional and non-functional requirements, whereas the solution space represents design elements \cite{AR7}. To create decision models for microservices systems, we represent the problem space as a set of QAs and the solution space as a set of microservices patterns and strategies. We developed the decision models for four microservices design areas, i.e., application decomposition, microservices security, microservices communication, and services discovery, because most of the design, development, and testing challenges are rooted in these areas (\cite{waseemMSAdevops, waseem2021nature, waseemMSAdesign}). The research method to conduct this study comprises three phases, each detailed below and illustrated in Figure \ref{fig:methodology}.

\subsection{Identifying MSA Patterns and Strategies}
\label{sec:RMethod}

\begin{table}[!t] \footnotesize \caption{Selected databases, search engines, and search string}
\begin{centering}
{\small{}}%
\begin{tabular}{|c|c|}
\hline 
\multicolumn{2}{|c|}{\textbf{\small{}Search string}}\tabularnewline
\hline 
\multicolumn{2}{|c|}{\emph{\small{}((microservice{*} OR micro service{*} OR  micro-service{*} }}\tabularnewline
\multicolumn{2}{|c|}{\emph{\small{}OR microservices architect{*} OR microservices design)}}\tabularnewline
\multicolumn{2}{|c|}{\emph{\small{}AND (pattern OR tactic OR quality attribute))}}\tabularnewline
\hline 
\textbf{\small{}Database} & \textbf{\small{}Targeted search area}\tabularnewline
\hline 
{\small{}ACM Digital Library} & {\small{}Paper title, abstract}\tabularnewline
\hline 
{\small{}IEEE Explore} & {\small{}Paper title, keywords, abstract}\tabularnewline
\hline 
{\small{}Springer Link} & {\small{}Paper title, abstract}\tabularnewline
\hline 
{\small{}Science Direct} & {\small{}Paper title, keywords, abstract}\tabularnewline
\hline 
{\small{}Wiley Online} & {\small{}Paper title, abstract}\tabularnewline
\hline 
{\small{}Engineering Village} & {\small{}Paper title, abstract}\tabularnewline
\hline 
{\small{}Web of Science} & {\small{}Paper title, keywords, abstract}\tabularnewline
\hline 
{\small{}Google Scholar} & {\small{}General search}\tabularnewline
\hline 
{\small{}Google} & {\small{}General search}\tabularnewline
\hline 
\end{tabular}{\small\par}
\par\end{centering}
\label{tab:stringDatabase}
\end{table}

We collected required patterns, strategies, QAs, and impact of patterns on QAs for creating decision models by reviewing scientific (e.g., journals and conference papers) and grey literature (e.g., blog posts and white papers) (see the Selected Studies sheet in the Replication Package \cite{replpack}). 
The following four steps \cite{Garousi2019} are used to extract the relevant studies and required data from both scientific and grey literature.

\textit{\textbf{Step 1} - Initial search}: We extracted the related studies by executing a search string on eight major databases and Google (see Table \ref{tab:stringDatabase}). The search yielded 2058 scientific and 52 grey literature. 
    
\textit{\textbf{Step 2} - Title and keywords based studies selection}: The extracted studies in scientific and grey literature were divided into two parts, and two authors (i.e., the first and sixth) further screened the literature independently by reading their titles and keywords. Both researchers excluded several hundred irrelevant scientific and grey literature that were not related to our study goal. Any uncertain literature during this screening process was discussed among all the authors to get a consensus. The title and keywords-based screening finally got 228 scientific and 25 grey literature. 

\textit{\textbf{Step 3} – Abstract and topic based studies selection}: During this step, the first author read the abstracts of the 228 scientific studies and labelled each study as “relevant”, “irrelevant”, or “doubtful”. The doubtful studies were discussed among all the authors for getting consensus about the relevance to our research context. On the other hand, the sixth author read the topics of the 25 grey literature and followed the same selection process. We finally selected 38 scientific and 22 grey literature based on their abstracts and topics (see the Selected Studies sheet in the Replication Package  \cite{replpack}).

\textit{\textbf{Step 4} - Data extraction and analysis}: We initially identified 211 patterns (strategies) from 39 scientific and 174 patterns (strategies) from 23 grey literature related to the four design areas, i.e., microservices decomposition, security, communication, and discovery. We found that studies use different names for the same pattern (strategy). We tried to understand each pattern (strategy) and identified the common naming for them. We also found that the terms “pattern” and “strategy” were used interchangeably. For instance, API rate limiting is a security pattern, but it is also discussed in the literature as a strategy (e.g., slow down attackers). After removing duplicate patterns and using common naming for several patterns, we finally got 7 patterns and strategies for application decomposition into microservices, 8 patterns and strategies for microservices security, 15 patterns and strategies for microservices communication, and 6 patterns and strategies for service discovery. Moreover, a particular pattern or strategy may have a positive or negative effect on QAs. During the data extraction and analysis, we also identified and analyzed the positive and negative impact of patterns and strategies on QAs (see the Patterns Impact sheet in the Replication Package \cite{replpack}).

\subsection{Modeling Decision Models}
\begin{figure}[!t]
 \centering
\includegraphics[width=8.5cm,height=8.5cm,keepaspectratio]{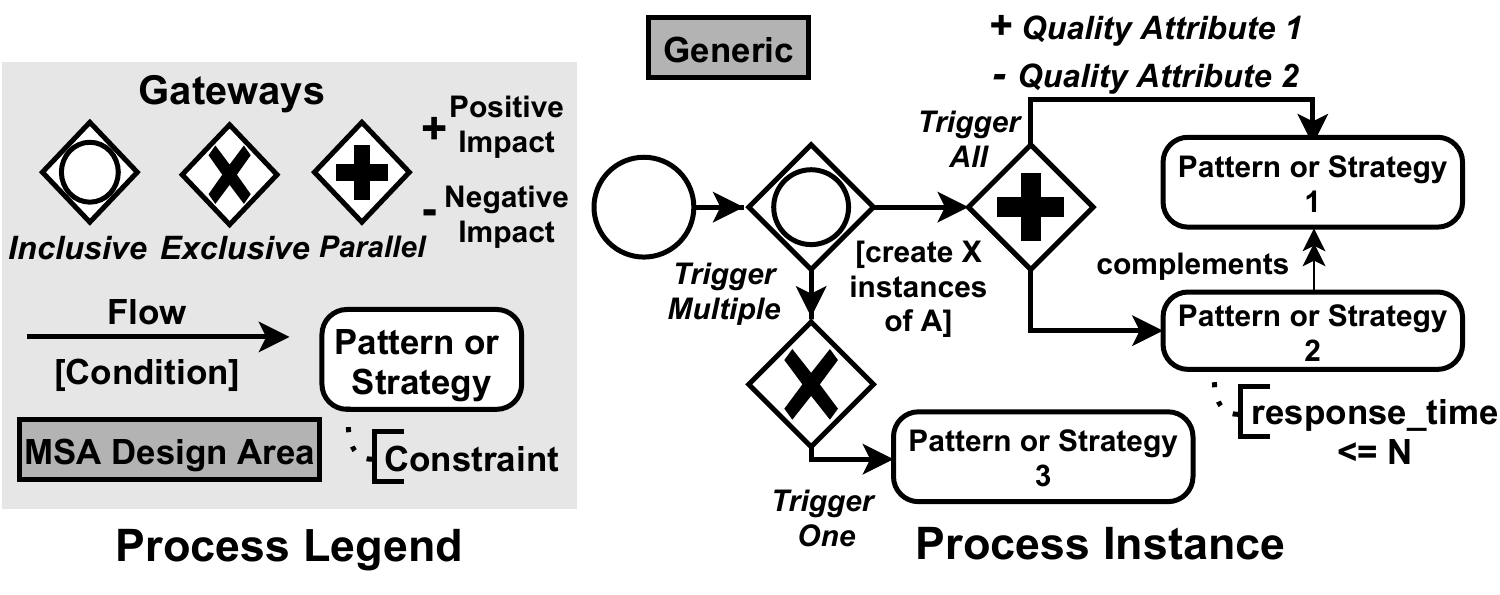}
\caption{Notations used in the decision models}
\label{fig:researchThemeProcess}
\end{figure}

Figure \ref{fig:researchThemeProcess} presents the notations used in the decision models presented in this paper. We used \textit{Inclusive}, \textit{Exclusive}, and \textit{Parallel} gateways of Business Process Model and Notation (BPMN) for indicating the decision flow. Each MSA design area is represented through \textit{grey box}. A \textit{circle} is used to denote the start of a decision process. An \textit{Inclusive gateway} is used to trigger more than one outgoing paths within a decision process. An \textit{Exclusive gateway} is used to trigger one of the outgoing paths. In contrast, A \textit{Parallel gateway} triggers all of the outgoing connected paths. We used \textit{rounded rectangle} to represent the patterns and strategies belong to an MSA design area. A \textit{double-headed} arrow shows a “complements” relationship between two patterns or strategies. An \textit{octagon and dashed} arrow is used to represent the constraints of each pattern or strategy. Finally, plus (+) and minus (-) signs indicate the negative impact of each pattern or strategy on the QAs.

\subsection{Evaluating Decision Models} 
To evaluate and refine the decision models, we conducted 24 semi-structured interviews with microservices practitioners from 24 medium and large companies from 12 countries. The interviewees were recruited through personal contacts, publicly available email addresses of microservices project contributors on GitHub, and social and professional platforms (e.g., microservices groups on LinkedIn, Facebook, and Google). 
We adapted the interview questions used in previous studies (e.g., \cite{AR6}, \cite{AR7}) to evaluate the decision models (see the Questionnaire sheet in the Replication Package \cite{replpack}). The interviews were conducted with a questionnaire which consists of five sections, including the questions about i) participant demographic information (5 questions), ii) decision model for application decomposition (8 questions), iii) decision model for microservices security (8 questions), iv) decision model for microservices communication (8 questions), and v) decision model for service discovery (8 questions). The interview questionnaire has both closed and open-ended questions.Participating in the interviews was voluntary with no compensation. The proposed models and interview questionnaire were shared with the interviewees 6 to 7 days before the interview.

\section{Decision Models}
\label{sec:modelsdescripation}

\subsection{Application Decomposition Decision Model}
\begin{figure}[!t]
 \centering
\includegraphics[scale=0.37]{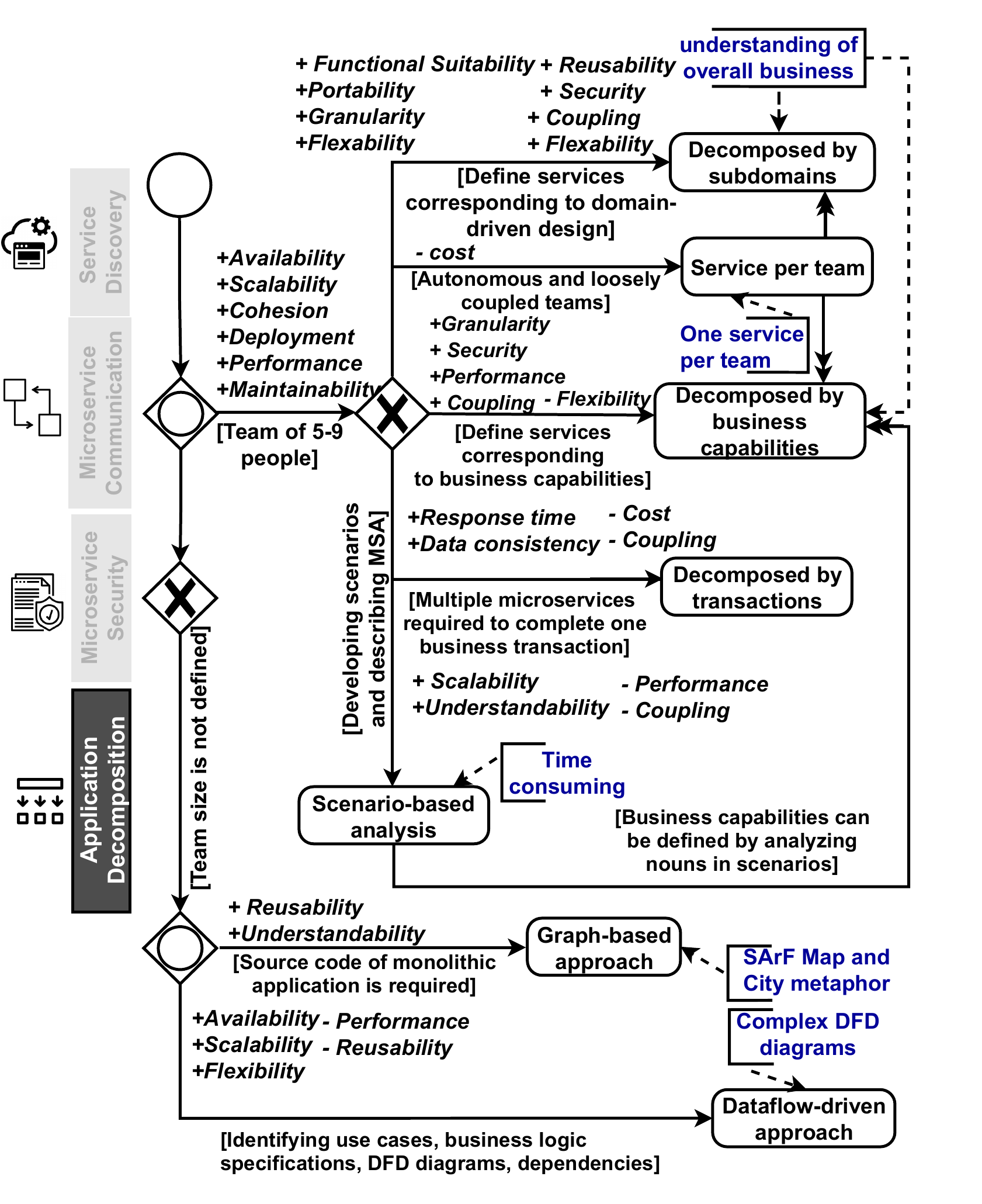}

\caption{Decision model for application decomposition}
\label{fig:AppDecopmositionModel}
\end{figure}

Monolithic applications need to be decomposed into small, independent, and loosely coupled microservices to achieve the benefits (e.g., improved scalability, independent deployment). Table \ref{tab:AppDecPat} lists the patterns and strategies covered by the application decomposition decision model (see Figure \ref{fig:AppDecopmositionModel}). The decision process for application decomposition into microservices is based on the team size and impact of patterns and strategies on QAs. If the application needs to be decomposed into microservices for the team of 5-9 people to increase \textit{Availability}, \textit{Scalability}, \textit{Cohesion}, \textit{Deployment}, \textit{Performance}, and \textit{Maintainability}, we can use one among seven illustrated patterns (see Figure \ref{fig:AppDecopmositionModel}). In the following, we further explain the other conditions, impact on QAs, and constraints for each pattern.

 To increase \textit{Flexibility}, \textit{Granularity}, \textit{Reliability}, \textit{Reusability}, \textit{Security}, \textit{Functional suitability}, and \textit {Portability}, \textbf{Decomposed by subdomains} pattern can be used. This pattern guides practitioners in defining each microservice responsibility, boundaries, and relationships with other microservices. To successfully implement this pattern, practitioners need to understand the overall business (see Figure \ref{fig:AppDecopmositionModel}). In contrast, if microservices need to be defined with respect to business capabilities,  \textbf{Decomposed by business capabilities} pattern can be used. Normally, business capabilities are organized into a multi-level hierarchy and generate business value. This pattern improves the \textit{Granularity}, \textit{Performance}, and \textit{Security} of microservices if the business capabilities are identified by understanding the client organization’s structure, purposes, and business processes. However, this pattern reduces \textit{Flexibility} because the application design is tightly coupled with the business model. Another option that we can use for decomposing applications is \textbf{Service per team} pattern. This pattern enables practitioners to break applications into microservices that individual teams can manage. It also complements \textbf{Decomposed by subdomains} and \textbf{Decomposed by business capabilities} patterns. A constraint of \textbf{Service per team} pattern is that only one small team (e.g., 5-9 people) owns one microservice, meaning that each team independently develops, tests, deploys, and scales individual microservice. The teams also interact with other teams to negotiate APIs. \textbf{Service per team} pattern increases \textit{Availability}, \textit{Scalability}, \textit{Cohesion}, \textit{Deployment}, and \textit{Performance}, and \textit{Maintainability}. If the project is large and needs to hire more people, \textbf{Service per team} pattern negatively impacts the development cost of microservices. 

Another exclusive pattern option in decomposition patterns is \textbf{Decompose by transactions}, in which applications are decomposed based on business transactions. Each business transaction carries one task, and each microservice has the functionalities for several business transactions (e.g., sales, marketing). This pattern allows grouping multiple microservices to avoid latency issues. \textbf{Decompose by transactions} pattern can help to improve \textit{Response time}, \textit{Data consistency}, and \textit{Availability} of microservices. Meanwhile, decomposing applications based on transactions also increases \textit{Execution cost} and \textit{Coupling} of microservices due to multiple functionalities being implemented in one microservice. Another option to decompose an application is \textbf {Scenario-based analysis} which consists of several steps, such as developing scenarios, describing MSA, and evaluating scenarios. During the evaluation process of scenarios, practitioners identify the microservices and interactions between them. This pattern is appropriate if practitioners have enough time to develop and describe the scenarios and MSA, respectively. This strategy can also be used to identify the business capabilities of systems by analyzing the nouns and verbs from given business scenarios. The identified nouns represent the microservices, and the verbs describe the relationship among them. While this strategy increases \textit{Scalability}, \textit{Performance} and \textit{Coupling} could be compromised due to the imprecise boundaries of microservices.

\begin{table}[!t]
    \footnotesize
    \caption{Application decomposition patterns and strategies}
    \begin{tabularx}{\columnwidth}{p{2.5cm}|X}
        \hline
        \textbf{Name} & \textbf{Summary}\\\hline
        Decomposed by subdomains \cite{richardson2018microservices} \cite{AWS} & Define services corresponding to Domain-Driven Design (DDD) subdomains.\\\hline
        Decomposed by business capabilities \cite{richardson2018microservices} \cite{AWS} & Define services corresponding to business capabilities.\\\hline
        Service per team \cite{richardson2018microservices} \cite{AWS} & Break down the application into microservices that individual teams can manage.\\\hline
        Decomposed by transactions \cite{AWS} & An application typically needs to call multiple microservices to complete one business transaction. To avoid latency issues, services can be defined based on business transactions.\\\hline
        Scenario analysis \cite{tusjunt2018refactoring} & Identify the business capabilities by analyzing the nouns and verbs from given business scenarios.\\\hline
        Graph-based approach \cite{kamimura2018extracting} & Identify microservices from the source code of existing monolithic applications by graph clustering and visualization techniques.\\\hline
        Data Flow-Driven (DFD) approach \cite{li2019dataflow} & Follow a top-down approach in which data flow diagrams contain the business requirements that are later partitioned through a formal algebra algorithm for identifying microservices.\\\hline
    \end{tabularx}
    \label{tab:AppDecPat}
\end{table}

Suppose that the team size is not defined for designing and developing microservices, and we need to identify the microservices from the code of legacy applications. In that case, we can choose \textbf{Graph-based approach}, which uses the SArF clustering algorithm to decompose the system for comprehension \cite{kobayashi2013sarf} and the city metaphor techniques for visualizing the system structure \cite{kamimura2018extracting}. The use of this approach increases the \textit{Reusability} of the existing code. \textbf{Graph-based approach} also visualizes the extracted microservices and their relationships along with the structure of the whole system. Hence, it also increases the \textit {Understandability} of the MSA design. Finally, if the team size is not defined and applications need to be decomposed by using DFDs, in that case, \textbf{Data flow-driven} approach can be used, which consists of several steps, such as eliciting and analyzing the business requirements for identifying use cases and business logic specifications, creating fine-grained DFDs, identifying the dependencies between processes and datastores, and identifying microservices by clustering processes and related data stores. \textbf{Data flow-driven} approach increases \textit{Availability}, \textit{Scalability}, and \textit{Flexibility}. In contrast, it decreases \textit{Performance} and \textit{Reusability} mainly due to complex DFDs. 

\subsection{Microservices Security Decision Model}
\begin{figure}[!t]
 \centering
\includegraphics[scale=0.37]{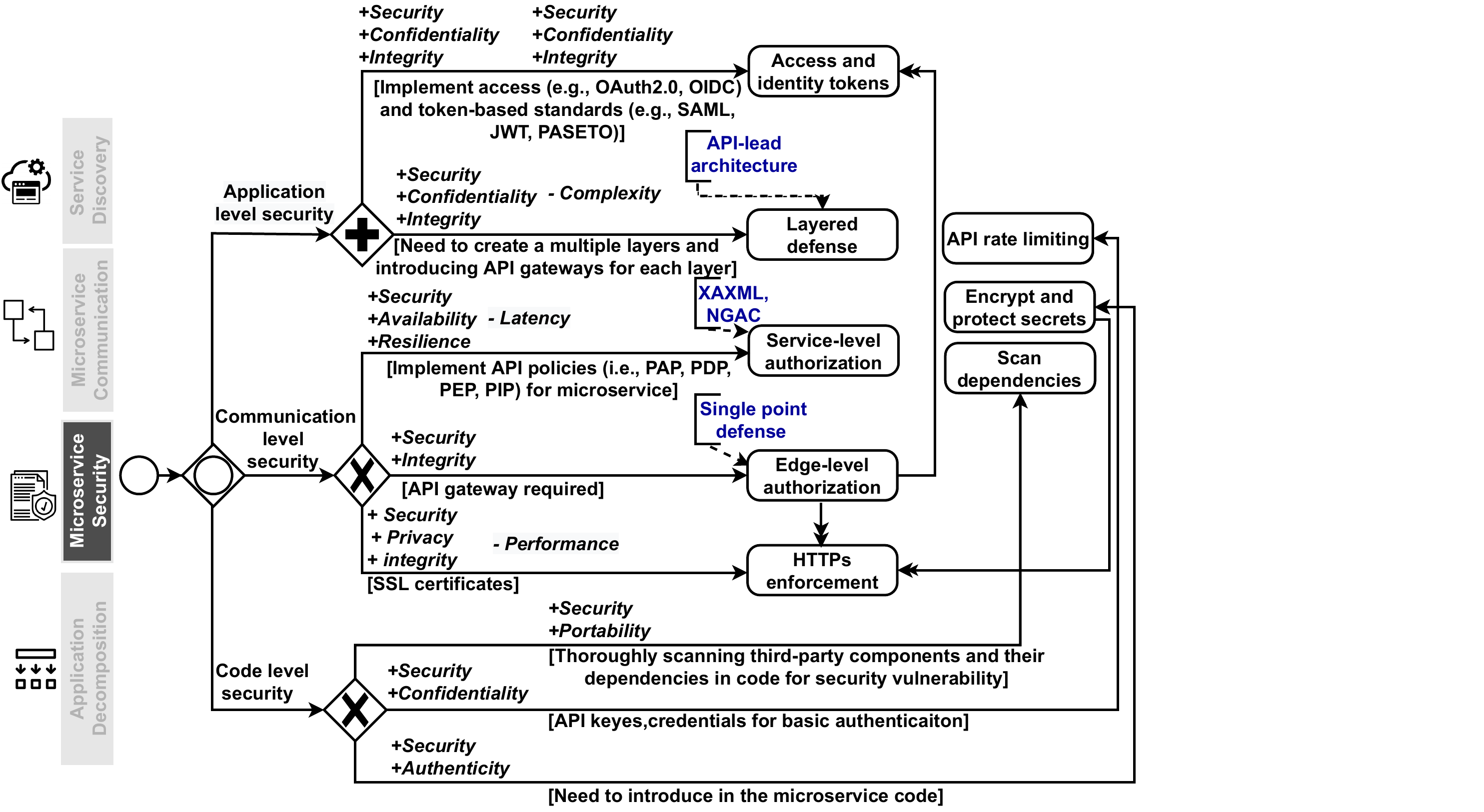}
\caption{Decision model for microservices security}
\label{fig:SecurityModel}
\end{figure}

The distributed nature of microservices systems makes them a potential target for cyber-attacks. Practitioners need to secure microservices systems at the application, communication, and code levels. Table \ref{tab:SecurityPatterns} lists the patterns and strategies covered by the microservices security decision model (see Figure \ref{fig:SecurityModel}). If microservices need to be secured at the application level to improve \textit{Confidentiality}, \textit{Integrity}, \textit{Accountability}, \textit{Authenticity}, and \textit{Recoverability}, \textbf{Access and identity tokens} pattern can be used, which encapsulates the security credentials of users (e.g., user identity, user group, user privilege) for accessing microservices through API gateways. It can be implemented by following several access-based and token-based standards, such as OAuth, OAuth2, OpenID, HTTP Basic Auth, and JSON Web Token (JWT), which enable microservices to verify that the requester is authorized to perform specific or all operations according to the given privilege. Similarly, \textbf{Layered defense} pattern could be used to secure microservices at the application level. \textbf{Layered defense} pattern provides the layered defense-in-depth for microservices systems, and it can be implemented by following the “API-led architecture” in which the whole application is converted into different APIs layers according to the functionality domain. Every API layer has a separate API gateway containing authentication and authorization policies specific to the API layer. This pattern increases \textit{Security}, \textit{Confidentiality}, and \textit{Integrity} because API gateways make it difficult for an intruder to penetrate deep into the system, while in the meantime, it increases \textit{Complexity} of microservices systems.

If practitioners need to secure microservices at the communication level, they can use \textbf{Service-level authorization} pattern, \textbf{Edge-level authorization} pattern, and \textbf{HTTPS enforcement} strategy. The \textbf{Service-level authorization} pattern gives more freedom to each microservice to control and enforce the access control policies for communication, which consist of several API policies, i.e., Policy Administration Point (PAP), Policy Decision Point (PDP), Policy Enforcement Point (PEP), and Policy Information Point (PIP). These access control policies are implemented through Extensible Access Control Markup Language (XACML) and Next Generation Access Control (NGAC) languages and notions. These API policies are mainly implemented in three ways \cite{OWASP}: i) directly implementing PDP and PEP at the microservices code level, ii) introducing a centralized policy repository containing a single policy decision point and implementing PDP and PEP at the microservices code level, and iii) introducing a centralized policy repository containing multiple policy decision points and implementing PDP and PEP at the microservices code level. This pattern improves \textit{Security}, \textit{Availability}, and \textit{Resilience}. However, it has a negative impact on \textit{Latency} because of additional network calls of the remote PDP endpoint. \textbf{Edge-level authorization} can also be used to secure microservices communication. This pattern enables the authorization at the edge level (API gateway). API gateway can consolidate authorization for all downstream microservices. Authorization at the edge level is hard to be implemented in a complex ecosystem due to many roles and access control policies. Moreover, because this pattern only secures API gateway, it violates the defense-in-depth policy. However, this pattern increases \textit{Security} and \textit{Integrity} of microservices systems. \textbf{Edge-level authorization} pattern can complement with \textbf{HTTPs enforcement} strategy that suggests the use of HTTPS connections instead of HTTP, and HTTPS implements a Secure Sockets Layer (SSL) protocol for establishing an encrypted link to secure the communication between microservices. 

The code of microservices can be secured by using \textbf{API rate limiting}, \textbf{ Encrypt and protect secrets}, and \textbf{Scan dependencies} strategies. \textbf{API rate limiting} strategy is used to slow down the attacks from intruders. The intruders use hundreds of gigs of the username and password combinations to breach security. This strategy can be implemented in microservices code or with an API gateway. This pattern not only increases the \textit{Security} and \textit{Authenticity} of microservices, but also protects microservices systems from abusive actions, including excessive API calls and rapidly updating configurations. \textbf{Encrypt and protect secrets} strategy is used to secure the microservices secretes (e.g., API key, user credentials). The secrets related to microservices can be stored using various key management services, such as Azure KeyVault, HashiCorp Vault, Spring Vault, and Amazon KMS. Finally, with \textbf{Scan dependencies} strategy, scanning programs (e.g., Dependabot) are used to scan the deployment pipeline, the primary line of code, the released code, and new code contribution from developers to detect security vulnerabilities. 

\begin{table}[!t]
\footnotesize
    \caption{Microservices security patterns and strategies}
    \begin{tabularx}{\columnwidth}{p{2.5cm}|X}
        \hline
        \textbf{Name} & \textbf{Summary}\\\hline
        Access and identity tokens \cite{SecurityPatterns1,SecurityPatterns2,richardson2018microservices} & 
       Verifies that a user is authorized to perform specific operations or not\\\hline
        Layered defence \cite{SecurityPatterns1} & Protect microservices systems by introducing multiple gateways and API-lead architecture\\\hline
        Service-level authorization \cite{OWASP} &  Give freedom to each microservice to control and enforce the access control policies for communication\\\hline
        Edge-level authorization \cite{OWASP} & Secure the edge points (API gateway) of microservices\\\hline
        HTTPS enforcement \cite{SecurityPatterns1,SecurityPatterns2,OWASP} & Suggests using HTTPS instead of HTTP to secure communication between microservices.\\\hline
        API rate limiting \cite{SecurityPatterns1, SecurityPatterns2}& Slow down the attacks from intruders\\\hline
        Encrypt and protect secrets \cite{SecurityPatterns1, SecurityPatterns2, SecurityOCTA} & Use tools (e.g., HashiCorp Vault, Microsoft Azure Key Vault, Amazon KMS) to secure the API key, user credentials, and other credentials related to microservices.\\\hline
        Scan dependencies \cite{SecurityPatterns1, SecurityPatterns2} & Scanning programs are used to detect the security vulnerabilities that may occurs because of dependency issues \\\hline
    
    \end{tabularx}
    \label{tab:SecurityPatterns}
\end{table}

\subsection{Microservices Communication Decision Model}
\begin{figure}[!t]
 \centering
\includegraphics[width=8.5cm,height=8.5cm,keepaspectratio]{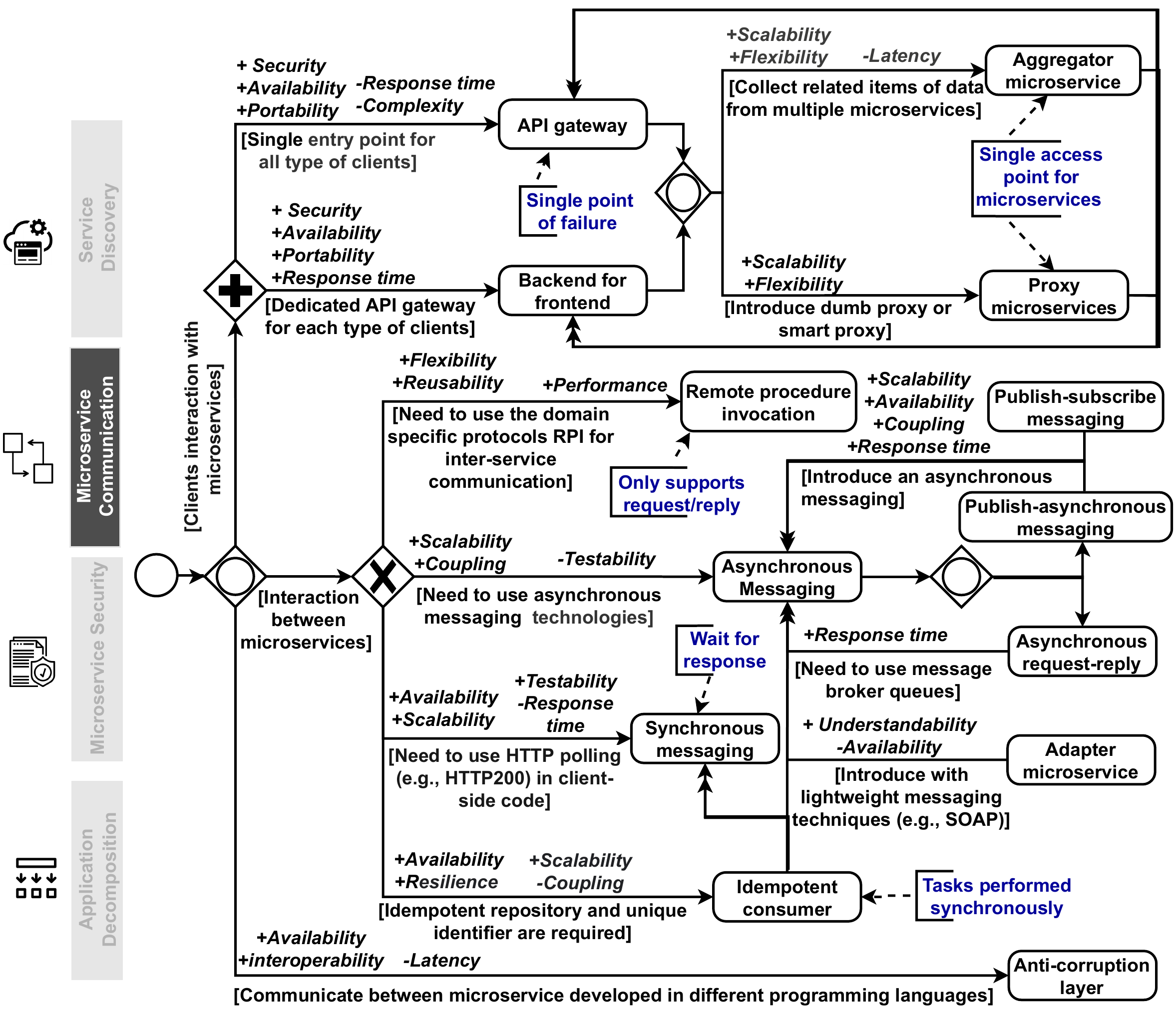}
\caption{Decision model for microservices communication}
\label{fig:CommunicationModel}
\end{figure}

Microservices systems consist of several independent services running on multiple servers or hosts. Each service or service instance communicates with other services using several patterns. Table \ref{tab:CommunicationPatterns} lists the patterns and strategies covered by the microservices communication decision model (see Figure \ref{fig:CommunicationModel}).

\begin{table}[!t]
\footnotesize
    \caption{Microservices Communication patterns and strategies}
    \begin{tabularx}{\columnwidth}{p{2.5cm}|X}
        \hline
        \textbf{Name} & \textbf{Summary}\\\hline
        API gateway \cite{richardson2018microservices}& Provide a single entry point to clients for accessing microservices\\\hline
       
        Backend for frontend \cite{richardson2018microservices} & Define a separate API gateway according to type of application client\\\hline
        
        Aggregator microservice \cite{2017Architectural} & Collect related items of data from multiple microservices\\\hline
        
        Proxy microservices \cite{2017Architectural} & Collect related items of data from multiple microservices through dumb and smart proxies\\\hline
        
        Remote procedure invocation \cite{richardson2018microservices}  & Establish inter-service communication via a request/reply-based protocol\\\hline
        
        Asynchronous messaging \cite{richards2015microservices} & Message sender does not wait for response of corresponding recipient microservices\\\hline
        
        Publish-subscribe messaging \cite{richards2015microservices,2017Architectural} & Allow sender microservice to broadcast the message to zero or more recipient microservices \\\hline
        
        Publish-asynchronous messaging \cite{richards2015microservices,2017Architectural} & Allow sender microservice to broadcast the message to one or more recipient microservices and get the response from some recipient microservices\\\hline
        
        Asynchronous request-reply \cite{richards2015microservices,2017Architectural} & Allow sender microservice to directly send a request message to a recipient microservice and get the immediate response \\\hline
        
        Synchronous messaging\cite{2017Architectural} & Message sender waits for response of corresponding recipient microservices\\\hline
        
        Idempotent consumer \cite{richardson2018microservices} & Detect and discard duplicate messages from sender microservices\\\hline
        Anti-corruption layer \cite{Satish} & Used to communicate the polyglot microservices \\\hline
    
    \end{tabularx}
    \label{tab:CommunicationPatterns}
\end{table}

The decision process for microservices communication begins with deciding how the application clients (mobile browsers, desktop browsers) interact with microservices. Usually, a single client needs to fetch data from multiple microservices to complete one transaction. The interaction between the application clients and microservices becomes possible by implementing \textbf{API gateway} and \textbf{Backend for frontend (BFF)} patterns. \textbf{API gateway} provides a single entry point for application clients to access microservices underneath. API gateway also improves \textit{Security}, \textit{Availability}, and \textit{Portability}. However, this pattern also increases \textit{Response time} because of the additional network hop and \textit{Complexity} due to the development, deployment, and management of API gateway. A variant of \textbf{API gateway} is \textbf{BFF} pattern. \textbf{BFF} pattern defines a separate API gateway (e.g., Web app API gateway, mobile API gateway, public API gateway) for each type of application clients (e.g., Web apps, mobile apps, 3rd party apps). This pattern is not appropriate for single interface (e.g., only Web interface) microservices systems. Similar to API gateway, \textbf{BFF} also improves \textit{Security}, \textit{Availability}, and \textit{Portability} in microservices systems. This pattern decreases \textit{Response time} due to dedicated API gateways for each type of application clients. We identified two other communication patterns that exclusively complement \textbf{API gateway} and \textbf{BFF} patterns. The first pattern is \textbf{Aggregator microservice} pattern that collects related items of data by invoking multiple microservices and returns them to the application clients via API gateway or BFF gateways (e.g., Web BFF API gateway, mobile BFF API gateway). \textbf{Aggregator microservice} pattern increases \textit{Scalability} and \textit{Flexibility}. However, it has a negative impact on \textit{Latency}. A variant of \textbf{Aggregator microservice} pattern is \textbf{Proxy microservices} pattern, in which different microservices can be invoked according to business needs. However, this pattern uses dumb and smart proxies. The dumb proxy only delegates the request to targeted microservices and returns the data to application clients without any data transformation. In contrast, a smart proxy applies data transformation before sending back the response to application clients \cite{2017Architectural}.

Inter-service communication often happens between microservices to complete business transactions. We identified several patterns that can be used for inter-service communication. \textbf{Remote Procedure Invocation (RPI)} enables inter-service communication between microservices through a request/reply-based domain-specific protocols (e.g., SMTP, IMAP, RTMP). Technologies like REST, gRPC, and Apache thrift can be used to implement \textbf{RPI} pattern. This pattern increases \textit{Flexibility}, \textit{Reusability}, and \textit{Performance}. The alternative patterns for inter-service communication are \textbf{Asynchronous messaging} and \textbf{Synchronous messaging}. Typically, \textbf{Asynchronous messaging} adopts a choreography style for inter-service communication, whereas \textbf{Synchronous messaging} adopts an orchestration style \cite{Katherine}. In \textbf{Asynchronous messaging} pattern, sender microservice does not wait for the response of corresponding recipient microservices. This pattern can be implemented by using several asynchronous messaging technologies, including Apache Kafka and RabbitMQ. \textbf{Asynchronous messaging} pattern has a positive impact on \textit{Scalability} and \textit{Coupling}. However, \textit{Testability} (debugging) for asynchronous messaging is difficult \cite{2018Microservice}. There are several patterns that are complemented with \textbf{Asynchronous messaging}, including \textbf{Publish-subscribe messaging}, \textbf{Publish-asynchronous messaging}, \textbf{Asynchronous request-reply}, and \textbf{Adapter microservice}. The conditions, QAs, and complements relations of these asynchronous messaging patterns are shown in Figure \ref{fig:CommunicationModel}. On the other hand, inter-service communication can also be possible through \textbf{Synchronous messaging} in which sender microservice waits for the response of corresponding recipient microservices. This pattern implements through HTTP calls. \textbf{Synchronous messaging} increases \textit{Availability}, \textit{Scalability}, \textit{Maintainability}, \textit{Testability}, and \textit{Coupling}. The \textit{Response time} of this pattern is relatively lower than \textit{Response time} of \textbf{Asynchronous messaging}. Furthermore, our decision model also contains \textbf{Idempotent consumer} pattern, which can be used with both \textbf{Asynchronous messaging} and \textbf{Synchronous messaging} patterns to handle duplicate messages for consumer services. This pattern detects and discards duplicates messages from sender microservices. Finally, \textbf{Anti-corruption layer} pattern can be used to communicate polyglot microservices. It increases \textit{Availability} and \textit{Interoperability}. However, \textit{Latency} is compromised due to an extra layer between microservices. \textbf{Anti-corruption layer} can also be used between legacy and new microservices systems for migration purposes.

\subsection{Service Discovery Decision Model}
\begin{figure}[!t]
 \centering
\includegraphics[width=8cm,height=8cm,keepaspectratio]{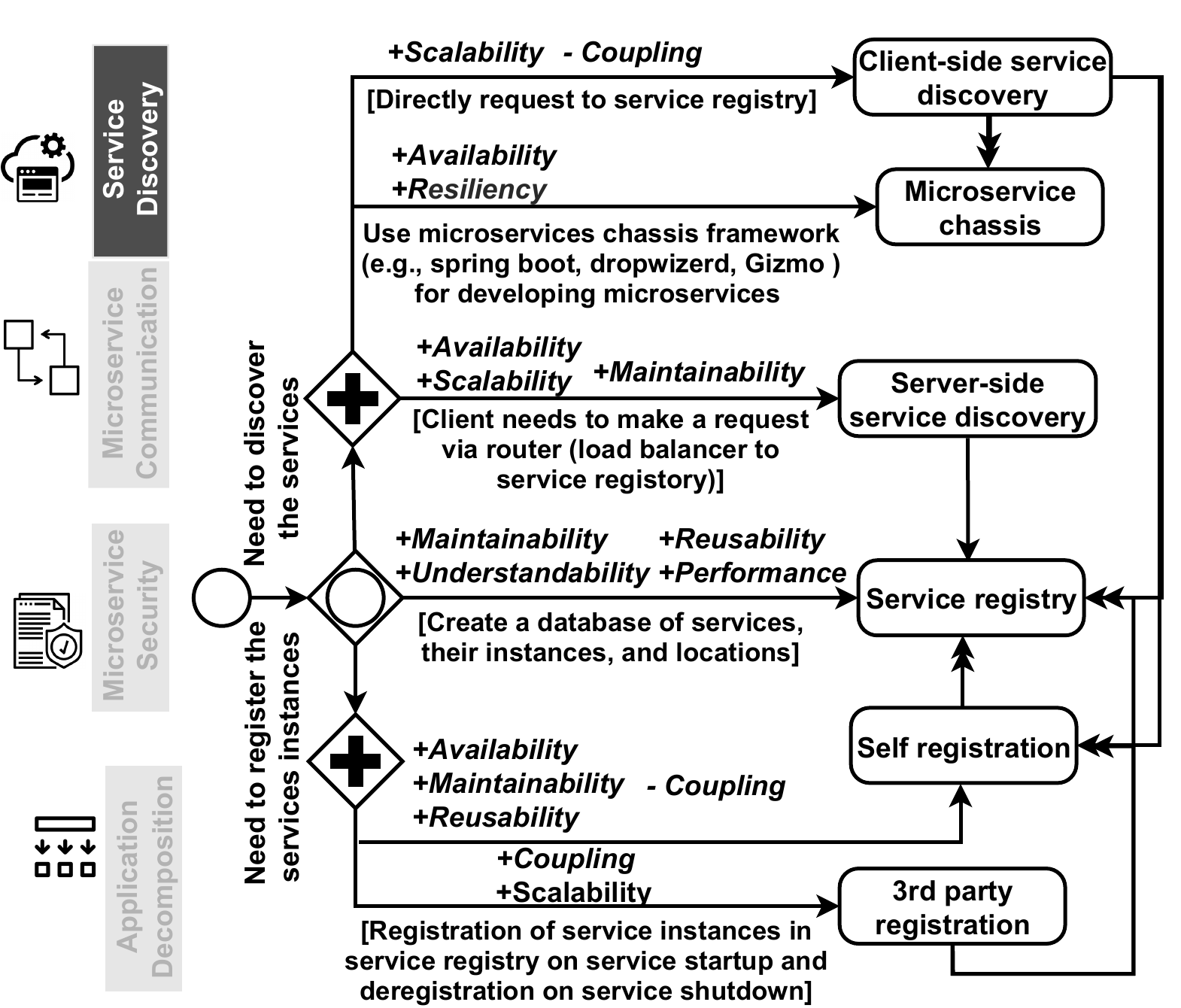}
\caption{Decision model for service discovery}
\label{fig:ServiceDiscovery}
\end{figure}

Microservices runs in a virtualized or containerized environment where the number of instances of a service and their locations dynamically change. A service’s client needs to discovery the latest location of the service instances for communication purpose. Table \ref{table:Servicediscovery} lists the patterns and strategies covered by the service discovery decision model (see Figure \ref{fig:ServiceDiscovery}). The patterns included in this model are integrated through a parallel gateway, meaning that all patterns can be implemented together. The central pattern of this decision model is \textbf{Services registry}, which is necessary for the implementation of all other service discovery patterns. Typically, each service has several instances which are hosted on virtual machines or containers with dynamic IP addresses. The number of instances increases or decreases according to the workload of the system, and IP addresses of instances change dynamically. For example, Amazon EC2 auto-scaling adjusts the number of instances according to the workload. \textbf{Service registry} pattern acts as a database of service instances and their locations.

At the first stage of the service discovery process, the service instances and their locations must be registered with service registry. The registration can be performed in two different ways, i.e., \textbf{Self registration} and \textbf{3rd party registration}. \textbf{Self registration} pattern enables service instances to register their hosts and IP addresses in service registry and makes themselves available for service discovery. Each service instance also needs to renew its registration with service registry periodically. This pattern improves the \textit{Scalability}, \textit{Maintainability}, and \textit{Reusability} of microservices. However, it also increases \textit{Coupling} because each service and its instances need to be registered with service registry. The alternative of \textbf{Self registration} is \textbf{3rd party registration pattern} that also registers service instances along with hosts and IP addresses in service registry when the services start and are unregistered when services shut down. This pattern increases \textit{Scalability}, whereas it decreases \textit{Coupling} between microservices.

If the clients of a service (e.g., API gateway) need to discover a service instance's current location, we can use \textbf{Client-side service discovery}, \textbf{Microservice chassis}, and \textbf{Server-side service discovery} patterns. \textbf{Client-side services discovery} pattern enables clients to directly request service registry for the location of the required service instances, and get a response. This pattern increases \textit{Scalability}. However, it also increases \textit{Coupling} due to direct calls between clients and service registry. To implement this pattern, we also need to implement separate service registration patterns according to the programming languages used to develop microservices. \textbf{Client-side service discovery} pattern complements \textbf{Self registration} and \textbf{Microservice chassis} patterns. \textbf{Client-side service discovery} is usually implemented with the help of \textbf{Microservice chassis} pattern. In \textbf{Microservice chassis} pattern, microservices are developed using different frameworks, such as Spring Boot, Spring Cloud, and Gizmo. \textbf{Microservice chassis} pattern also improves the \textit{Availability} and \textit{Resiliency} of microservices. Another alternative for service discovery is \textbf{Server-side service discovery} pattern, in which clients make a request via a router (i.e., load balancer) to service registry for the location of the required service instances, and get a response through the router. The implementation of \textbf{Server-side service discovery} is simpler than \textbf{Client-side service discovery} because it only makes the request to a router and the router requests service registry for the location of the service instances. 

\begin{table}[!t]
\footnotesize
    \caption{Service discovery patterns and strategies}
    \begin{tabularx}{\columnwidth}{p{2.5cm}|X}
        \hline
        Service registry \cite{richardson2018microservices,SDPMSA}& Hold the dynamic IP addresses of all service instances\\\hline
        Client-side service discovery \cite{richardson2018microservices,SDPMSA}& Directly access the dynamic addresses of service instances from service registry  \\\hline
        Server-side service discovery \cite{richardson2018microservices,SDPMSA} & Access the dynamic addresses of service instances via routers from service registry\\\hline
        Microservice chassis \cite{richardson2018microservices,SDPMSA} & Enable the implementation of client-side service pattern via Microservices chassis frameworks\\\hline
        Self registration \cite{richardson2018microservices,SDPMSA} & Enables microservices to register their instances with service registry on service startup and update service status periodically \\\hline
        3rd party registration \cite{richardson2018microservices,SDPMSA} & 3rd party registration pattern is an alternative solution of Self registration pattern\\\hline
    \end{tabularx}
    \label{table:Servicediscovery}
\end{table}

\section{Evaluation}
\label{sec:evaluation}
\begin{figure}[!t]
 \centering
\includegraphics[width=8.5cm,height=8.5cm,keepaspectratio]{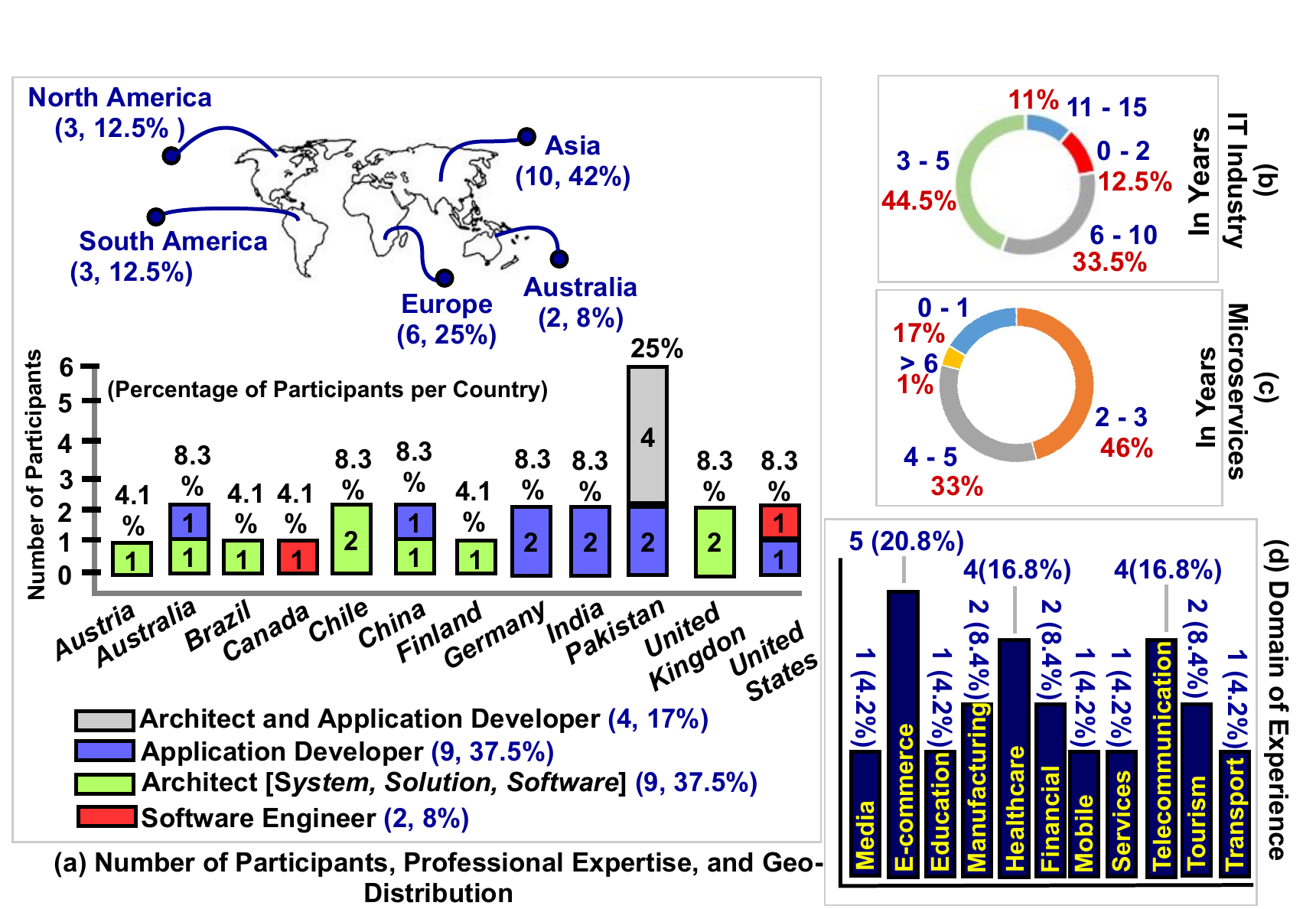}
\caption{Demography details of interviewed practitioners}
\label{fig:Intervieweesdemographics}
\end{figure}

To evaluate the decision models, we conducted 24 semi-structured interviews with a questionnaire \cite{replpack}. The key results of the evaluation are presented below:

\textbf{Demographics of interviewees}: 
The 24 interviewees (P1 to P24) come from 24 IT companies from 12 countries (see Figure \ref{fig:Intervieweesdemographics}-a). The roles of the participants are mainly related to the design and development of microservices systems (see Figure \ref{fig:Intervieweesdemographics}-a), such as application developer (9 interviewees), architect (9 interviewees), architect and application developer (4 interviewees), and software engineer (2 interviewees). 
11 interviewees have 2-3 years of experience, 8 interviewees have 4-5 years of experience,  4 interviewee has 0-1 year of experience, and only one interviewee has more than 6 years of experience working with microservices systems (see Figure \ref{fig:Intervieweesdemographics}-c). The domains of the participants’ organizations are mainly related to E-commerce, healthcare, financial services, and tourism (see Figure \ref{fig:Intervieweesdemographics}-d).

\begin{figure*}[!htbp]
  \centering
  \includegraphics[width=0.86\textwidth]{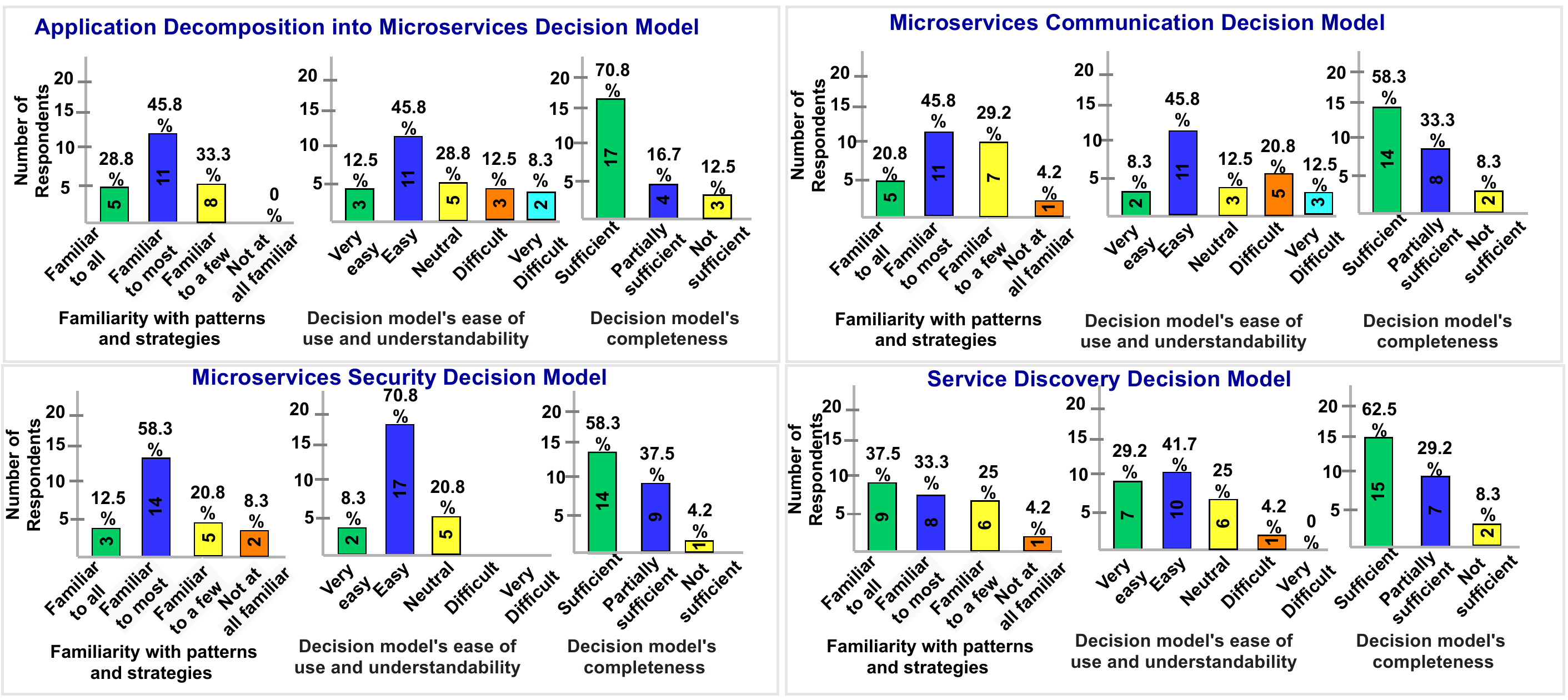}
  \caption{Overview of practitioners' responses for familiarity, understandability, and completeness of the decision models}
  \label{fig:Likertanswers}
\end{figure*}

\textbf{Familiarity with patterns and strategies}: We asked the interviewees, “\textit{Are you familiar with the patterns and strategies used to propose the decision models?}”. The majority of the interviewees mentioned that they are familiar with most of the patterns and strategies used to propose the decision models (see Figure \ref{fig:Likertanswers}). For instance, 11 (45.8\%) out of 24 interviewees mentioned that they are “familiar to most” of the application decomposition patterns and strategies, and 14 (58.3\%) interviewees mentioned that they are “familiar to most” of the security patterns and strategies. Only two (8.3\%) interviewees mentioned that they are not familiar with security patterns and strategies.

\textbf{Understandability and correctness of decision models}: Regarding the understandability of each decision model, we asked the interviewees, “\textit{To what extent decision models are easy to understand and use?}”. The interviews show that most of the models are easy to understand and easy to follow because of the self-explanatory flow (see Figure \ref{fig:Likertanswers}). For example, 17 (70.8\%) interviewees responded that the security decision model is easy to understand and use. Regarding completeness of the decision models, we asked the interviewees, “\textit{Does the information in this decision model sufficiently support making decisions about application decomposition, security, communication, and service discovery?}”. The interviews responded that most of the models are complete and sufficiently support decisions making (see Figure \ref{fig:Likertanswers}). For example, 17 (70.8\%) interviewees responded that the application decomposition into the microservices decision model sufficiently supports decision making. We also asked the interviewees, “\textit{Are these decision models correct? If not correct, please indicate the problem(s)}”. In response to this question, the interviewees indicated several minor issues, such as missing conditions, unclear complements relation, duplicate patterns and strategies, and incorrect impact of patterns on QAs. We fixed these issues in the updated version of the models.

\textbf{Usefulness for MSA design and development}: Regarding usefulness we asked,  “\textit{Are the decision models useful to support making decisions in the design and development of microservices systems? Why?}”. All the interviewees mentioned that every decision model sufficiently supports the corresponding MSA design area. Following is a representative answer about the usefulness of the microservices communication decision model.

\faComment{“\textit{This decision model illustrates several patterns and strategies for different kinds of communication styles, and each decision point also reflects the positive and negative impact of the patterns on QAs. Practitioners can easily decide which pattern or strategy they can apply to achieve the communication goal}”} \textbf{\textit{Architect and Application Developer (P6)}}.

\textbf{Usefulness for MSA evaluation}: We also asked the interviewees, “\textit{Are the decision models useful in the architecture evaluation of microservices systems? Why?}”. Most interviewees thought that the proposed decision models are useful in microservices architecture evaluation due to the availability of the constraints on patterns, conditions that need to be met by the patterns, trade-off on QAs, and impact of patterns and strategies on QAs. Following is a representative answer about the usefulness of the decision models in microservices architecture evaluation.

\faComment{“\textit{This set of decision models is helpful in evaluating different aspects of the microservices architecture. Each model proposes a number of design choices for the respective design area along with QAs that can be used for MSA evaluation}”} \textbf{\textit{Architect and Application Developer (P4)}}.

\textbf{Suggestions for improving decision models}: Finally, we asked the interviewees to provide their suggestions for improving the decision models. We received several suggestions in which the interviewees suggested presenting patterns with code, measure quantitative values for QAs, and use decision models in industrial microservices projects.

\section{Threats to Validity}
\label{sec:threats}





\textbf{Internal validity} helps to measure the soundness of conducted research, i.e., the extent to which data and evidence support the claims about decision models facilitating the architecting process for microservices systems. We addressed the following threats related to internal validity. \textit{Correctness of decision models}: We tried to mitigate this threat through collaborative work between the authors of this study and the analysis of practitioners' feedback. Regarding the collaborative work, two authors in the research team focused on identifying the decision models. In contrast, others used the information to visually represent and cross-check the models (see Step 1 and Step 2 in Figure \ref{fig:methodology}). On the other hand, we also considered the recommendations provided in the semi-structured interviews to improve the decision models. \textit{Responses of semi-structured interviews}: A potential threat in semi-structured interviews is subjectivity and human bias in answers to specific questions after using the models. Based on their experience, the types and industrial domains of the systems that the interviewed practitioners work with may impact their perspective towards design and architectural artifacts. Some of the interviewees may not reveal their true opinions about the decision models. To mitigate this threat, we conducted two pilot interviews to identify any issues, provided briefings and clarifications to practitioners before the interviews, and followed up with any clarifications or information during and after the interview.  Moreover, we presented illustrative examples of each decision model in order not to misinterpret the semi-structured interview questions.

The potential threats to \textbf{external validity} are related to the degree in which the results of a study can be generalized. In this respect, we considered the validation of decision models as a threat. Although the evaluation we conducted is based on a limited number of practitioners involved, however, we tried our best to ensure diversity in terms of distributed geo-locations, years of experience, type of professional roles, and industrial domains (see Figure \ref{fig:Intervieweesdemographics}) for rigorous evaluation. We believe that despite a limited number of interviews, practitioners' feedback helps to validate the models in terms of their familiarity, understandability, completeness, and usefulness in developing microservices systems. Still, we admit that our findings may not generalize and represent all microservices practitioners’ perspectives.

\textbf{Construct validity} is related to taking correct operational measures for collecting the data in this study. One potential threat is the inadequate use of the primary constructs of the decision models (i.e., MSA patterns and strategies, QAs, impact of the patterns on QAs). To mitigate this threat, we adopted the following operational measures: (i) we conducted a pilot search to ensure the correctness and appropriateness of the search terms, (ii) we used eight databases (see Table \ref{tab:stringDatabase}) in software engineering research for retrieving the scientific studies, and (iii) we used Google for searching the grey literature. Additionally, we followed the guidelines \cite{Garousi2019} to review and extract data from the scientific and grey literature.

The threats to \textbf{conclusion validity} affect the ability to reach correct conclusions.
In order to mitigate this threat, we defined a research methodology based on the practices and guidelines used in recent studies (e.g., \cite{AR6}, \cite{AR7}, \cite{Garousi2019}) to identify MSA patterns and strategies and to create and evaluate our decision models. Additionally, to ensure the reliability of our study, we have also made the Replication Package available \cite{replpack}.

\section{Related Work}
\label{sec:relatedWork}

\subsection{Decision Models and Guidelines for Architecting Microservices Systems}
During microservices system development, decision models \cite{AR2, haselbock2017decision} and practitioners’ feedback \cite{AR1, AR5} provide a set of guidelines (e.g., architectural models, patterns, recommended practices) that can empower the practitioners (e.g., architects) to rely on reusable knowledge and best practices to design, develop, validate, and evolve microservices systems \cite{AR7}.

\textbf{Decision guidance models for microservices systems}: Decision guidance models represent concentrated knowledge and rationalization about design decisions, such as modelling notations, patterns, and reference architectures to architect and develop microservices systems \cite {AR5}. The study in \cite{haselbock2017decision} examines existing literature and provides guidance models for microservices discovery and fault tolerance. Regarding migrating microservices systems, Ayas et al. \cite{ayas2021facing} identified three decision-making processes in microservices migrations consisting of 22 decision points and their alternative options by interviewing 19 participants. Data management aspects of microservice systems are reported in \cite{AR4}. Specifically, this study reports decision guidance models about generating, processing, and managing monitoring data, and disseminating monitoring data to stakeholders for the process automation domain. Harms et al. \cite{AR2} provide guidelines that support architects while selecting suitable front-end architecture(s) for microservices systems.

\textbf{Practitioners’ perspectives and recommendations for architecting microservices systems}: In contrast to decision models, practitioners’ perspectives (i.e., developers’ feedback) can streamline the industrial practices and experts’ recommendations for the design and development of microservices systems. Ntentos et al. \cite{AR1} present best practices and patterns for microservice systems. Based on identified practices and patterns, the authors have derived a formal architecture decision model containing 325 elements and relations. The derived architectural decision model reduces the (i) efforts needed to understand the architectural decisions for microservice data management and (ii) uncertainty in the design process. An empirical study in \cite{AR5} interviewed 10 microservices experts to identify 20 design areas and investigate (i) which design areas are relevant for microservices, (ii) how important they are, and (iii) why they are important.

\subsection{Decision Models for Selecting Patterns}
During the software development life cycle, selecting the most appropriate patterns that can be applied to a particular design context is a critical challenge. Haselb{\"o}ck et al. \cite{AR5} proposed a decision model that assists developers and architects in selecting appropriate patterns for blockchain-based applications. The proposed decision model was evaluated based on expert opinions regarding its correctness and usefulness in guiding the architecture design and understanding the rationale of various design decisions. The study \cite{AR7} presents decision models for cyber-foraging systems that maps functional and non-functional requirements to architectural tactics for cyber-foraging. 

\subsection{Conclusive Summary}
Reviewing related research suggests that there is a lack of research on decision models that can leverage patterns and strategies as reusable knowledge to address specific design areas of microservices systems (i.e., application decomposition, microservices security, communication, and service discovery). Existing research on pattern-based architecting of microservices systems \cite{haselbock2017decision} exploits decision models but lacks strategies associated with patterns and their impacts on quality attributes (i.e., architecturally significant requirements). In contrast to survey-based studies on finding suitability and application of decision models \cite{AR1}, our work first derives four decision models that can leverage a multitude of patterns and strategies addressing various MSA design areas, and then validates these decision models based on practitioners' feedback.
\section{Conclusions}
\label{sec:conclusions}

This research presents decision models that leverage patterns, strategies, and QAs for rationalizing design knowledge to assist architects and developers in engineering microservices systems. The decision models have been identified by systematically reviewing multivocal literature that advocates the role of patterns, architectural strategies, and QAs to support decomposition, security, communication, and service discovery aspects of microservices systems. To validate the decisions models, we engaged a total of 24 microservices practitioners to evaluate the familiarity, understandability, completeness, and usefulness of the decision models. From the practitioners' perspective, the proposed decision models, their underlying patterns, and strategies can empower MSA architects and developers to exploit reusable knowledge and recurring practices for architecture-centric development of microservices systems.

Future research aims to focus on (1) extending the proposed decision models (establishing a decision model repository), exploring the possible combination of patterns and strategies, and validating the models through an industrial scale case study, and (2) developing a recommendation system that supports automation and human-decision in selecting the most appropriate patterns and strategies while analyzing their impacts on quality attributes.


\section*{Acknowledgments} \label{sec:ack}
This work was funded by the National Key R\&D Program of China with No. 2018YFB1402800 and the NSFC with No. 62172311.
\balance

\bibliographystyle{ACM-Reference-Format}
\bibliography{main}

\end{document}